\documentclass[letter]{aa} 

\newcommand{\esothanks}{Based on observations collected at the
  European Organisation for Astronomical Research in the Southern
  Hemisphere under ESO programme 096.B-0690.}

\usepackage{graphicx}
\usepackage{txfonts}

\begin{document} 

\title{VLT/MUSE Illuminates Possible Channels for Lyman Continuum Escape in the
  Halo of SBS 0335-52E\thanks{\esothanks}}

\titlerunning{Ionised gas tails in the halo of SBS 0335-52E}
\authorrunning{E.~C.~Herenz et al.}

\author{
  E.~C.~Herenz\inst{\ref{inst1}}
    \and
  M.~Hayes\inst{\ref{inst1}}
    \and
  P.~Papaderos\inst{\ref{inst3}}
    \and
  J.~M.~Cannon\inst{\ref{inst2}}
    \and
  A.~Bik\inst{\ref{inst1}}
    \and
  J.~Melinder\inst{\ref{inst1}}  
    \and
  G.~Östlin\inst{\ref{inst1}}
}

\institute{
  Department of Astronomy, Stockholm University, AlbaNova
  University Centre, SE-106 91, Stockholm, Sweden \label{inst1}
  \and
  Instituto de Astrof\'{i}sica e Ci\^{e}ncias do Espaço - Centro de
  Astrof\'isica da Universidade do Porto, Rua das Estrelas, 4150-762
  Porto, Portugal\label{inst3} 
  \and 
  Department of Physics \&
  Astronomy, Macalester College, 1600 Grand Avenue, Saint Paul, MN
  55105, USA \label{inst2} }

\date{Received ...; accepted ...}
 
\abstract{We report on the discovery of ionised gas filaments in the
  circum-galactic halo of the extremely metal-poor compact starburst
  SBS\,0335-052E in a 1.5\,h integration with the MUSE integral-field
  spectrograph. We detect these features in H$\alpha$ and
  [\ion{O}{iii}] emission down to a limiting surface-brightness of
  $5 \times 10^{-19}$\,erg\,s$^{-1}$cm$^{-2}$arcsec$^{-2}$ .  The
  filaments have projected diameters of 2.1\,kpc and extend more than
  9\,kpc to the north and north-west from the main stellar body.  We
  also detect extended nebular \ion{He}{ii} $\lambda$4686 emission
  that brightens towards the north-west at the rim of a star-burst
  driven super-shell.  We also present a velocity field of the ionised
  gas.  The filaments appear to connect seamlessly in velocity space
  to the kinematical disturbances caused by the shell.  Similar to
  high-$z$ star-forming galaxies, the ionised gas in this galaxy is
  dispersion dominated.  We argue that the filaments were created via
  feedback from the starburst and that these ionised structures in the
  halo may act as escape channels for Lyman continuum radiation in
  this gas-rich system.  }


\maketitle

\section{Introduction}
\label{sec:introduction}

Pinpointing the sources that are relevant for reionising the Universe
at $z\gtrsim 7$ is one of the current major goals in observational
cosmology.  Hydrogen ionising photons ($E_\gamma \geq 1$\,Ry) from
low-mass star-forming galaxies are one viable candidate
\citep[e.g.,][]{Faisst2016,Schaerer2016}, but direct
observations of these photons leaking from high-$z$ galaxies are
scarce \citep[e.g.][]{Rutkowski2017}.  Moreover, the processes
enabling the escape of this so-called Lyman continuum (LyC) radiation
from the interstellar- and circum-galactic medium (CGM) are not well
constrained.  In principle feedback from supernovae can carve ionised
gas cavities through the CGM, thereby creating channels through which
the LyC can leak into the intergalactic medium \citep[][]{Fujita2003}.
Observational inferences of such processes have to be made from the
neutral and the ionisied gas phase in the CGM \citep{Bik2015}.  At high redshifts,
however, direct observations of the CGM in its neutral phase are
currently impossible \citep{Obreschkow2011} while directly detecting its
ionised phase remains challenging
\citep{Rauch2016,Wisotzki2015,Finley2017}.  Fortunately, observations
of nearby galaxies with properties comparable to systems at the
highest redshifts provide a promising alternative for detailed
examinations of the relevant physical processes in the early universe
\citep[for a recent review see][]{Hayes2015}.

The SBS\,0335-052 system at a redshift of $z=0.0135$ ($d=54$\,Mpc)
consists of a pair of extremely metal-deficient star-forming dwarf
galaxies separated by 22\,kpc (in projection) that has long been
recognised as a special laboratory for such studies
\citep[][]{Izotov1990}.  In fact, with
$12 + \log(\mathrm{O/H}) \lesssim 7.15$ (3\% of the solar value) the
western galaxy -- SBS\,0335-052W (discovered by
\citealt{Pustilnik1997}) -- is one of the most metal-poor emission
line galaxy known \citep[][]{Izotov2009}.  Observations of galaxies
with such low oxygen abundance are rare and only a handful of objects
comparable to SBS\,0335-052W are known
\citep[][]{Guseva2017,Izotov2017}.  The brighter eastern galaxy --
SBS\,0335-052E -- shows only a slightly higher nebular oxygen
abundance, and thus also belongs to the elusive group of extremely
metal-poor compact star-forming galaxies.  HST observations of this
galaxy reveal six young ($t<10 - 25$\,Myr) super-star clusters (SSCs, with
masses $M_\mathrm{SSC} > 10^5$M$_\odot$\ and star-formation rates
$\mathrm{SFR}_\mathrm{SSC} \lesssim 1$\,M$_\odot$yr$^{-1}$) that are
within $\sim500$\,pc of each other \citep{Thuan1997,Adamo2010}.  Observations of the
21cm line revealed that the two galaxies are surrounded by a large HI
complex \citep{Pustilnik2001}.  This HI halo exhibits elongated tidal
tails to the east and to the west as well as a faint diffuse bridge
connecting both galaxies \citep{Ekta2009}.  The inferred neutral
hydrogen column density towards brightest star-forming regions in
SBS\,0335-052E is $2\times10^{21}$\,cm$^{-2}$.  Analysis of
  HST UV spectroscopy with GHRS \citep{Thuan1997a} and COS
  \citep{James2014} points at an even higher neutral column of
  $5 - 7 \times 10^{21}$\,cm$^{-2}$.  Thus, the
optical depth for photons above 1\,Ry is $10^4$ or higher, implying
total absorption of LyC radiation from the young stellar population in
SBS\,0335-052E along our sightline. \\ \indent In this letter we
report on a discovery of low surface-brightness (SB) filamentary tails
in H$\alpha$ and [\ion{O}{iii}] emission that emanate to the north (N)
and north-west (NW) of SBS\,0335-052E.  We speculate that these highly
ionised tails may act as channels for LyC photon leakage.

\begin{figure*}
  \centering
  \includegraphics[width=0.49\textwidth]{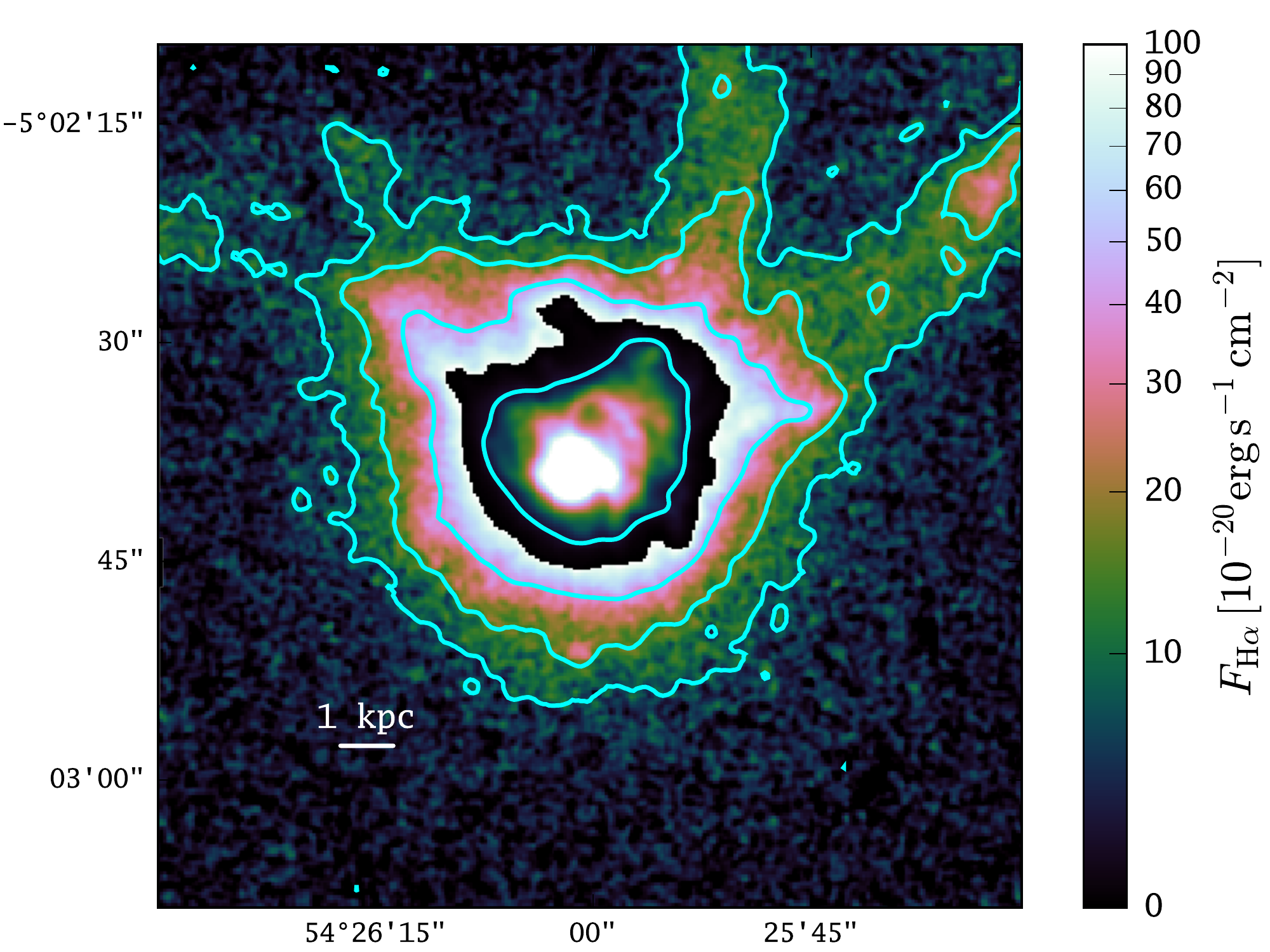}
  \includegraphics[width=0.49\textwidth]{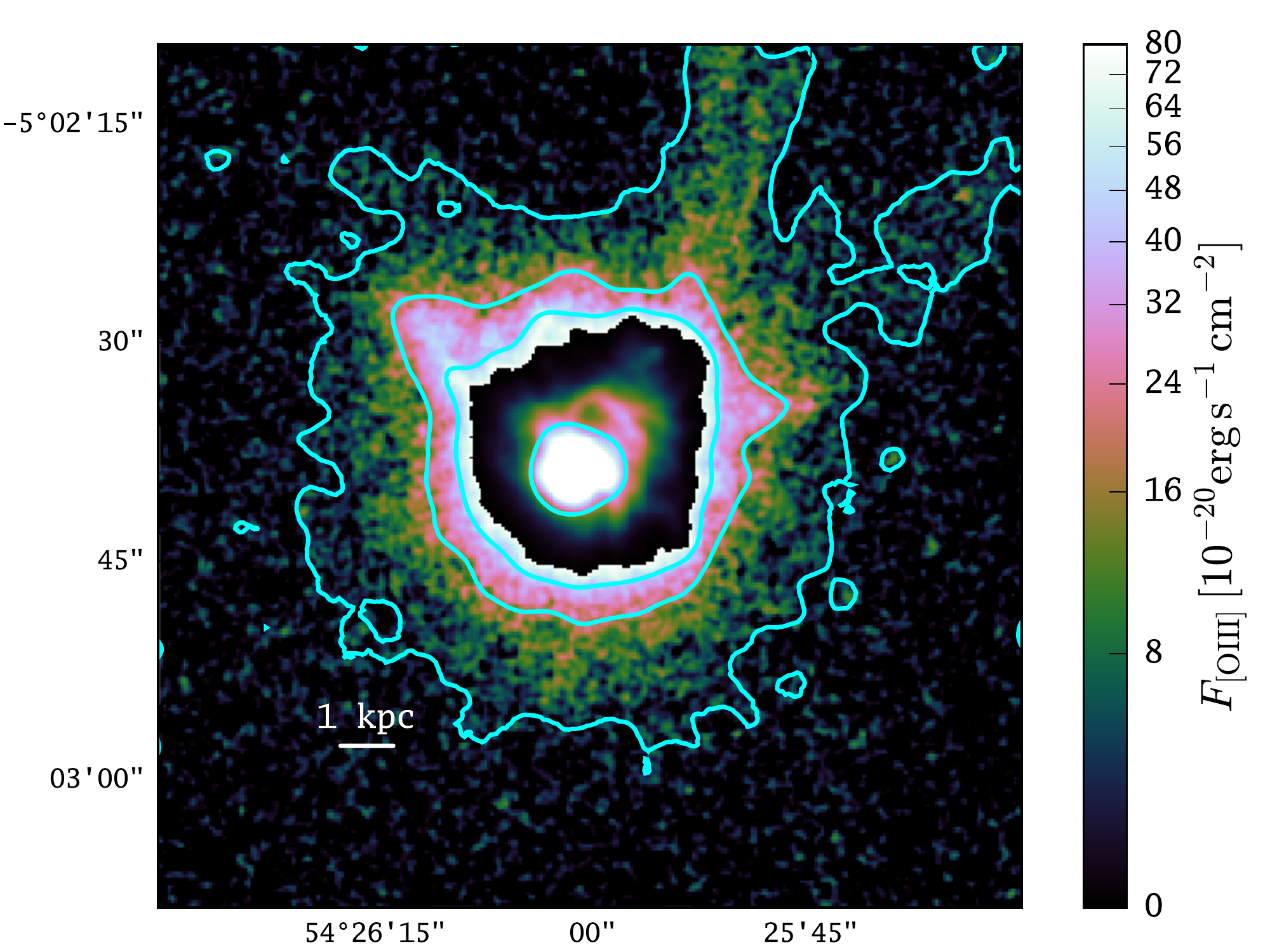}
  \caption{H$\alpha$ (\emph{left}) and [\ion{O}{iii}] $\lambda5007$
    (\emph{right}) narrow band images of SBS\,0335-052E created from
    the MUSE datacube.  East is left and North is up.  The colour
    scale within the high-SB region encodes fluxes from 10$^{-18}$ to
    $10^{-15}$\,erg\,s$^{-1}$cm$^{-2}$ for H$\alpha$ and
    $8\times10^{-19}$ to $8\times10^{-16}$\,erg\,s$^{-1}$cm$^{-2}$ for
    [\ion{O}{iii}], while for the outer low surface-brightness region
    the encoding is shown by the colour bar on the right. An
    asinh-scaling is used. SB contours are drawn at
    $[2.5,5,12.5,1250]\times
    10^{-18}$\,erg\,s$^{-1}$cm$^{-2}$arcsec$^{-2}$ for H$\alpha$ and
    $[0.5,5,12.5,1000]\times
    10^{-18}$\,erg\,s$^{-1}$cm$^{-2}$arcsec$^{-2}$ for
    [\ion{O}{iii}]. To highlight the low-SB features the images have
    been smoothed with a $\sigma=1$\,px (0.2\arcsec{}) Gaussian
    kernel. }
  \label{fig:narrow}
\end{figure*}
\begin{figure}
  \vspace{-0.8em}
  \centering
  \includegraphics[width=0.5\textwidth]{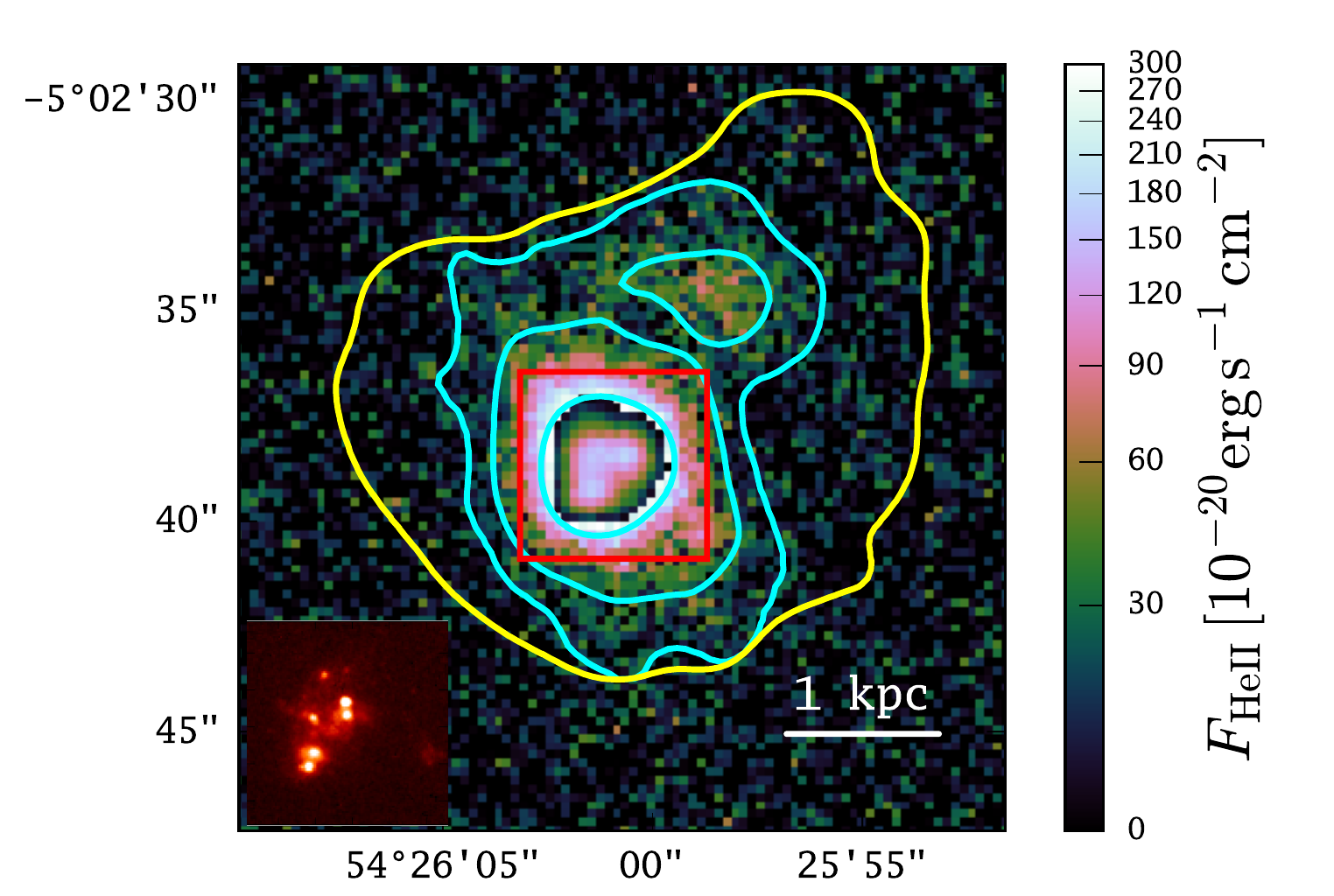} 
  \caption{\ion{He}{ii} narrow band image magnified on the central
    region.  The yellow contour is the
    $1.25\times10^{-15}$erg\,s$^{-1}$cm$^{-2}$arcsec$^{-2}$ contour from the
    H$\alpha$ image (Fig.~\ref{fig:narrow}, left).  Colour scale in
    the inner high-SB region extends from $3\times 10^{-18}$ to
    $3\times 10^{-17}$erg\,s$^{-1}$cm$^{-2}$arcsec$^{-2}$, while for
    the outer region it is shown by the colour bar on the right. An
    asinh-scaling is used. Contours correspond to
    $[2.5,100,1000]\times
    10^{-18}$\,erg\,s$^{-1}$cm$^{-2}$arcsec$^{-2}$. The inset in the
    bottom left shows the emsission line free HST WFC3/F550M image
    from the galaxy within the region indicated by the red square. }
  \label{fig:heiinb}
\end{figure}
\begin{figure}
  \vspace{-0.8em}
  \centering
  \includegraphics[width=0.454\textwidth]{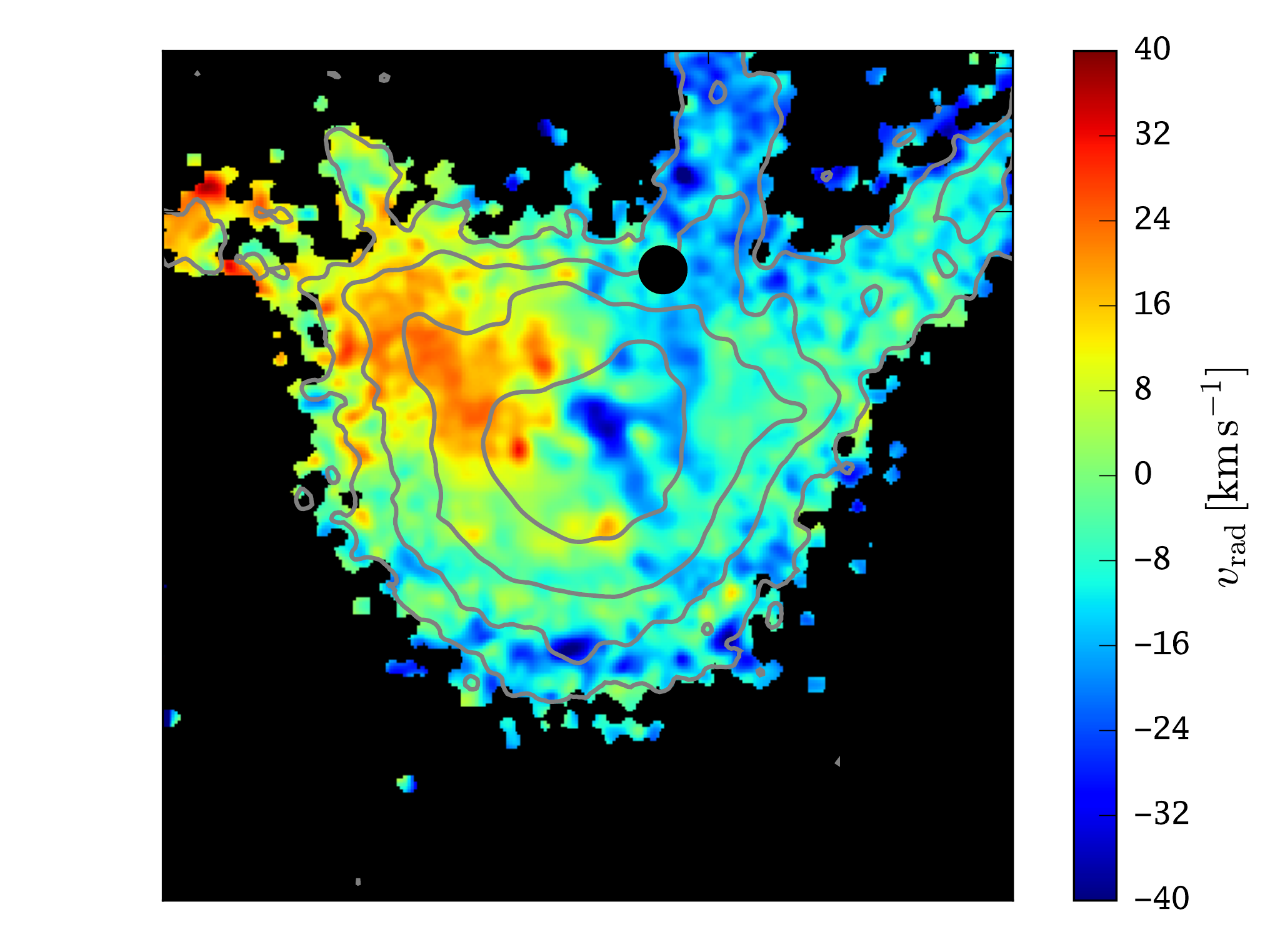}
  \vspace{-0.3em}
  \caption{H$\alpha$ line of sight velocity field with H$\alpha$ SB
    contours from Fig.~\ref{fig:narrow}. A background [\ion{O}{ii}]
    $\lambda$3727 emitter at $z=0.78$ coincident with H$\alpha$
    contaminated the fit towards the north and hence the respective
    region is masked (black circle). .  We spatially smoothed the
    datacube with a tophat filter ($r=1$\arcsec{}) to enhance the
    signal to noise for the fit.}
  \label{fig:vfield}
\end{figure}

\vspace{-1em}
\section{Observations, data reduction, and creation of narrow-band
  images }
\label{sec:observ-data-reduct}

VLT/MUSE \citep{Bacon2010} observations of SBS\,0335-052E were
obtained in service mode under clear skies on November 16th and 17th,
2015.  The seeing varied between 0.9\arcsec{} and 1.2\arcsec{}.  We
used the wide-field mode without the second-order blocking filter. The
total exposure time was 5680\,s (8$\times$710\,s integrations).  No
separate sky exposures were taken.  We reduced and calibrated the data
using the MUSE data reduction system \citep[DRS][]{Weilbacher2014} version 1.6.2.  After
reduction we were equipped with a sky-subtracted, wavelength, and flux
calibrated datacube covering a field of view (FoV) of 1\arcmin{}$\times$1\arcmin{} over
a spectral range from 4600\AA{}--9370\AA{}.  From this cube we
produced an ``emission-line'' only datacube by subtracting an in
spectral direction median-filtered (180\AA{} filter width)
``continuum-only'' cube \citep[see Sect.~4.1 in][]{Herenz2017}.  We then
synthesised 8\AA{} wide narrow-band images centred around the emission
lines H$\alpha$, [\ion{O}{iii}] $\lambda$5007, and the \ion{He}{ii}
$\lambda$4686, thus probing zones ionised by $E_\gamma \geq 1$Ry,
$E_\gamma \geq 2.58$Ry and $E_\gamma \geq 4$Ry photons, respectively.
The final images are shown in Figure~\ref{fig:narrow} for H$\alpha$
and [\ion{O}{iii}] (after applying the correction described below) and
Figure~\ref{fig:heiinb} for \ion{He}{ii}.

Due to the large spatial extent of the H$\alpha$ and [\ion{O}{iii}]
emission the standard sky correction resulted in over subtraction of
H$\alpha$ and [\ion{O}{iii}] flux.  This resulted in negative
zero-point offsets in the H$\alpha$ and [\ion{O}{iii}] narrow-band
images and, moreover, visible H$\alpha$ and [\ion{O}{iii}] emission
line features in the DRS calculated sky spectrum.  After ensuring that
no real skylines were present at the affected wavelengths, we removed
those features from the calculated sky spectrum by linear
interpolation over the bracketing wavelengths.  Feeding this modified
spectrum back into the DRS removed the over subtraction for H$\alpha$
completely, while in the [\ion{O}{iii}] band a remaining small average
offset of $2\times10^{-20}$\,erg\,s$^{-1}$cm$^{-2}$ had to be
corrected manually.

\vspace{-1em}
\section{Results}
\label{sec:results}

In the high SB region of our H$\alpha$ and [\ion{O}{iii}] images
(Fig.~\ref{fig:narrow}) we resolve in great detail the super-shell
structure($r \approx 1.3$\,kpc) first discussed in
\cite{Thuan1997}. Extending further to the N and NW from that shell
two low-SB filaments are seen.  In the north-east (NE) a third, less
pronounced filament is also noticeable.  A deep
  long-slit spectrum by \cite{Izotov2001} already indicated the
  presence of extended emission in the SE - NW direction.  Our new
  MUSE data is significantly deeper, detecting extended H$\alpha$ as
  faint as $5\times 10^{-19}$\,erg\,s$^{-1}$cm$^{-2}$arcsec$^{-2}$.
  Moreover, the spatially resolved nature of our MUSE data allows to
  map the filamentary morphology of the extended low-SB emission for
  the first time.  The filaments appear to continue beyond the MUSE
FoV, permitting us only to provide lower limits of their radial extend:
 9.7\,kpc (35.1\arcsec{}) and 9.3\,kpc (33.7\arcsec{}) for
the NW and N filament, respectively.

The filamentary ``ears'' differ in their [\ion{O}{iii}]/H$\alpha$ flux
ratios.  The NW ear is only barely detected in [\ion{O}{iii}], while
it brightens in H$\alpha$ towards the edges of the FoV, so that
[\ion{O}{iii}]/H$\alpha \lesssim 1/10$.  The northern ear, however, is
clearly detected in [\ion{O}{iii}], with
[\ion{O}{iii}]/H$\alpha \approx 1/4$.  These values are much lower
compared to the central high-SB region. There (i.e., within the
1.25$\times 10^{-15}$erg\,s$^{-1}$cm$^{-2}$arcsec$^{-2}$ H$\alpha$ SB
contour) we measure [\ion{O}{iii}]/H$\alpha = 0.99$ and
[\ion{O}{iii}]/H$\beta = 2.97$.  The [\ion{O}{iii}]/H$\beta$ (or
[\ion{O}{iii}]/H$\alpha$) ratio depends on metallicity and on the
ionisation parameter (i.e. the ratio of ionising photon density to gas
density).  The lower ratio found in the ears could be due to lower
density and/or lower metallicity in the CGM.  Unfortunately, other
than only marginal detections in [\ion{O}{iii}] $\lambda$4959 and
H$\beta$, we do not find other nebular lines from the filaments in our
data.  Hence, we can not make strong inferences on the physical
conditions in the ears.  Even H$\beta$ is not detected significantly
enough to permit the calculation of H$\alpha$/H$\beta$ ratios in those
low-SB regions that would be needed to asses the presence of dust.

Compared to its main stellar body, SBS\,0335-052E also exhibits
significantly extended \ion{He}{ii} $\lambda$4686 emission
(Figure~\ref{fig:heiinb}).  The distinct peaks in the high-SB
\ion{He}{ii} image are co-spatial with the more evolved SSCs,
indicating that here the \ion{He}{ii} emission is related to post-main
sequence stars (Wolf-Rayet stars) and/or their remnants
\citep[as first obtained by][]{Izotov2006}.  Moreover, towards the
NW the \ion{He}{ii} emission is more elongated with a brightening on
the rim of the shell.  This points towards a locally enhanced UV
radiation field caused by the shell shockfront
\citep{Izotov2001,Thuan2005}.  The total \ion{He}{ii} flux, obtained
by integrating over all spaxels within the
$2.5\times 10^{-18}$\,erg\,s$^{-1}$cm$^{-2}$arcsec$^{-2}$ SB contour,
is
$F_\mathrm{HeII} = (2.17 \pm 0.03) \times
10^{-15}$erg\,s$^{-1}$cm$^{-2}$.  Other possible sources for producing
\ion{He}{ii} ionising $E_\gamma > 4$\,Ry photons are very massive
metal-poor O-stars or high-mass X-ray binaries
\citep{Thuan2004,Prestwich2013}.

We investigate the kinematics of the ionised gas by fitting a 1D
Gaussian profile to the H$\alpha$ line in each spaxel. The resulting
line of sight velocity field is shown in Figure~\ref{fig:vfield}.
Within the high-SB region our velocity field agrees with the results
and \cite{Moiseev2015} who traced H$\alpha$ emission to
$\sim$10$^{-17}$\,erg\,s$^{-1}$cm$^{-2}$arcsec$^{-2}$.  These authors
interpreted the central blue-shifted arc-like features as kinematic
disturbances caused by the expansion of the super-shell.  The shell
structure manifests itself also in a locally double peaked H$\alpha$
profile, which is however only resolved in the VLT/GIRAFFE data from
\cite{Izotov2006} that was taken at $\sim 5 \times$ higher spectral
resolution compared to MUSE.  The newly detected filaments connect
seamlessly in velocity space to the central velocity field.
Especially, the ear towards the north shows the same blueshift as the
shell. The extended kinematical imprint of the shell in the filaments
argues for them being created by feedback from the starburst.  To
quantify the ionised gas kinematics we calculate the shearing velocity
$v_\mathrm{shear}$ and the intrinsic velocity dispersion\footnote{We
  correct for instrumental broadening using the wavelength dependent
  width of the MUSE line spread function given in Fig.~5 of
  \cite{Husser2016}.} $\sigma_0$ \citep[see review
by][]{Glazebrook2013}.  Following the procedures detailed in
\cite{Herenz2016} we obtain $v_\mathrm{shear} = 19.8$\,km\,s$^{-1}$
and $\sigma_0 = 29.2$\,km\,s$^{-1}$. Our derived $\sigma_0$ is in
excellent agreement with the measurement by \cite{Moiseev2015}:
$\sigma_0 = 30.6 \pm 1.6$\,km\,s$^{-1}$.  As
$v_\mathrm{shear} / \sigma_0 = 0.68 < 1$, SBS\,0335-052E qualifies as
a dispersion dominated system.  Dispersion dominated kinematics are
commonly found among high-redshift star-forming galaxies
\citep[e.g.,][]{Turner2017}. \cite{Herenz2016} showed that
$v_\mathrm{shear} / \sigma_0 < 1$ appears to be a necessary, but not
solely sufficient, criterion for galaxies to have Lyman $\alpha$
photons escaping.  The escape of Lyman $\alpha$ photons can be
theoretically linked to the leakage of LyC emission
(\citealt{Verhamme2015}, for recent observational evidence see
\citealt{Verhamme2017}).  However, towards our line of sight
Ly$\alpha$ is strongly absorbed \citep{Thuan1997a,James2014}, and
there are only tentative signatures of a possible extended low-SB
Ly$\alpha$ halo \citep{Ostlin2009}.  However, as we will argue in
Sect.~\ref{sec:summary-discussion}, we believe that the low neutral
column along the filaments will facilitate LyC (and possibly
Ly$\alpha$) escape in this galaxy.

Lastly, we also point out that the galaxy is embedded in diffuse
extended H$\alpha$ emission.  Detections of diffuse ionised halo gas
far away from the main stellar body have been presented for individual
spiral galaxies \citep[e.g.,][]{Hlavacek-Larrondo2011} and in stacks
of SDSS spectra \citep{Zhang2016}.  Measurements of this diffuse
component could provide constraints on the multi-phase nature of the
CGM.  We claim a significant detection out to 15\arcsec{} (4.1\,kpc)
from the main stellar body at
$\sim10^{-18}$\,erg\,s$^{-1}$cm$^{-2}$arcsec$^{-2}$ (lowest iso-SB
contour in Figure~\ref{fig:narrow}).  Constructing a SB profile
without including the regions of the ears, we even measure diffuse
emission out to 30\arcsec{} at
$\gtrsim 10^{-19}$\,erg\,s$^{-1}$cm$^{-2}$arcsec$^{-2}$.  However,
given the compactness of the high-SB region -- 95\% of the total
encircled energy are within 2\arcsec{} -- and that the SB-profile
spans more than six orders of magnitude within 30\arcsec{}, diffuse
scattered light from telescope optics and/or the atmosphere can not be
neglected as a cause for the observed halo \citep[][]{Sandin2014}.  As
detailed by \cite{Sandin2014} an accurate determination of the MUSE
point-spread out to radii $\gtrsim30$\arcsec{} would be needed to
accurately account for this effect, but such a measurement is not
available.

\vspace{-1em}
\section{Discussion and Conclusion}
\label{sec:summary-discussion}

Star formation in SBS\,0335-052E is concentrated in six SSCs
\citep{Thuan1997} that have ages $\lesssim 10$\,Myr and masses
$\sim 5\times 10^{5} - 10^{6}$\,M$_\odot$.  The two oldest ($\sim 7$
-- $10$\,Myr), most massive ($\sim 10$ -- $20 \times 10^5$\,M$_\odot$)
SSCs are in the north and the two youngest ($\lesssim 3$\,Myr), less
massive ones ($\sim 4$ -- $5\times 10^5$\,M$_\odot$) are in the south
\citep{Thompson2009,Adamo2010}.  The star-burst might have been
triggered either by the recent encounter with its western companion or
by a weak tidal interaction with the massive spiral NGC 1376
($M\sim10^{12}$\,M$_\odot$), which is found at a projected distance of
150\,kpc \citep{Pustilnik2001}.  The SSCs show an age-sequence within
490 pc (1.9\arcsec{}) from NW to south-east (SE).  The age gradient
points towards a scenario where feedback from the oldest cluster in
the NW ignited subsequent star-formation (SF) in the SE regions.  This
picture of unidirectional propagating SF as the main driver of the
buildup of SBS\,0335-052E is strengthened by an age gradient in faint
stellar clusters detected with an unsharp-masking technique out to
$\gtrsim$1 kpc NE of the brightest SSC \citep{Papaderos1998}.

In the opposite direction of SF propagation, i.e. towards the
lower density regions in the NW, feedback from the starburst is
responsible for the super-shell structure.  Initially the surface of
such shells are very dense and hence optically thick for ionising
photons.  However, with continued mechanical energy input from SF the
shells get accelerated and, due to the density contrast with the
ambient halo gas, Rayleigh-Taylor instabilities form.  These fragment
the surface of the shell and thus open funnels through which ionising
photons start leaking into the halo.  The precise timeline of bubble
evolution depends sensitively on the density distribution of the gas and
the mechanical energy input from the starburst \citep{Fujita2003}, but
since we observe the ionised filaments the shell must be past this
so-called ``blow-out''.  Ionising photons are now able to leak
through the ``cracks'' of the shell into the CGM of much lower density
thereby creating the observed ionised filaments.  Moreover, the hot
wind that pushes through the cracks also contributes in ionising the
ambient medium via shocks \citep{Cooper2008}.

Further inferences can be made when comparing our MUSE observations to
inteferometric 21cm imaging of the system.  The highest-spatial
resolution \ion{H}{I} images ($\sim$4\arcsec{} beam) by
\cite{Ekta2009} show indeed that the highest column densities
($N_\mathrm{HI} \sim 10^{21}$\,cm$^{-2}$) are confined to the
south-east, i.e. opposite the direction of the expanding shell.  With
increased sensitivity at lower spatial resolution ($\sim$7\arcsec{}
beam), \ion{H}{i} at lower column density ($\sim 10^{20}$\,cm$^{-2}$)
aligned with the direction of the shell becomes visible.  Both these
observational facts are consistent with the invoked scenario of
feedback driven SF propagation in the high-density SE regions, and the
feedback driven compression of neutral gas at the shell front towards
the regions of lower density in the NE.  At even lower resolution and
higher sensitivity the images by \cite{Ekta2009} reveal a tidal
feature slightly west of the NW ionised filament with
$N_\mathrm{HI} \approx 7\times 10^{19}$\,cm$^{-2}$.  While the spatial
resolution at this sensitivity appears insufficient to infer a direct
spatial connection between filaments and neutral halo gas, it still
shows that the bulk of $N_\mathrm{HI} \gtrsim 10^{19}$\,cm$^{-2}$ gas
must be within the MUSE FoV, and that both the N and NW filament
extend into regions of lower neutral columns.  Thus, while we do not
cover the full extend of the filaments, they appear to have
penetrated completely through the bulk of the
neutral gaseous halo of the system.  Consequently, the reduced neutral
fraction within those ionised filaments could advocate an escape of
LyC photons from the starburst into the intergalactic medium.

To conclude, we have presented arguments for the ionised filaments of
SBS\,0335-052E towards the N and NW being caused by SF-driven
feedback, while feedback driven SF propagation proceeds towards the
opposite direction in this galaxy.  Interestingly, a large fraction of
extremely metal-deficient star-burst galaxies show cometary
morphologies.  This is regarded as evidence for unidirectional SF
propagation being common in such systems \citep{Papaderos2008}.
Moreover, 7\%-10\% of high-$z$ galaxies show similar morphologies
\citep{Straughn2006}.  In order to explain cometary morphologies (also
known as ``tadpole'' galaxies) also other mechanisms have been put
forth, namely early-stage mergers or stream-driven gas accretion from
the cosmic-web
\citep[e.g.,][]{Rhoads2005,SanchezAlmeida2014,Straughn2015,Lagos2016}.
Nevertheless, we speculate that if propagating SF is indeed an
important driver of dwarf galaxy formation at high redshift, a highly
anistropic LyC escape into small solid angles would be expected.  In
this case only a small fraction of LyC leakers will be directly
detectable.  More importantly, accounting for this effect might be
important when budgeting the ionising photon output from low-mass
star-forming galaxies in order to asses their contribution for
reionising the universe.

\begin{acknowledgements}
  We thank P.~M.~Weilbacher for feedback on handling the
  sky-oversubtraction issue described in
  Sect.~\ref{sec:observ-data-reduct}.  M.H. acknowledges the support
  of the Swedish Research Council, Vetenskapsr\aa{}det and the Swedish
  National Space Board (SNSB), and is Fellow of the Knut and Alice
  Wallenberg Foundation.  This work was supported by Funda\c{c}\~{a}o
  para a Ci\^{e}ncia e a Tecnologia (FCT) through national funds and
  by FEDER through COMPETE by the grants UID/FIS/04434/2013 \&
  POCI-01-0145-FEDER-007672 and PTDC/FIS-AST/3214/2012 \&
  FCOMP-01-0124-FEDER-029170.  PP was supported by FCT through
  Investigador FCT contract IF/01220/2013/CP1191/CT0002.
\end{acknowledgements}

\vspace{-2em}

\bibliographystyle{aa}
\bibliography{sbs0335_letter_SUBMISSION.bib}

\end{document}